\documentclass[twocolumn,showpacs,aps]{revtex4}

\usepackage{graphicx}  
\usepackage{dcolumn} 
\usepackage{bm} 

\input psfig.sty

\input rotate 

\def\bea{\begin{eqnarray}}
\def\eea{\end{eqnarray}}
\def\ben{\begin{equation}}
\def\een{\end{equation}}
\def\benu{\begin{enumerate}}
\def\enu{\end{enumerate}}

\def\n{n}

\def\sss{\scriptscriptstyle\rm}

\def\1var{(\bx_1...\bx\N)}

\def\half{\frac{1}{2}}

\def\bx{{x}}

\def\x{_{\sss X}}
\def\c{_{\sss C}}
\def\s{_{\sss S}}
\def\xc{_{\sss XC}}

\def\N{_{\sss N}}
\def\H{_{\sss H}}

\def\ext{_{\rm ext}}

\def\sph_int{ {\int d^3 r}}

\def\PRA{Phys. Rev. A\ }
\def\PRB{Phys. Rev. B\ }
\def\PRL{Phys. Rev. Letts.\ }
\def\JCP{J. Chem. Phys.\ }

\newcommand{\arccoth}{\;\mbox{arccoth}\;}

\newcommand{\sech}{\;\mbox{sech}\;}
\newcommand{\csch}{\;\mbox{csch}\;}
\newcommand{\h}{\hat{H}}

\begin{document} 
\headheight 50pt

\preprint{RUTGERS DFT GROUP: pre-print}

\title{Density functional theory in one-dimension for contact-interacting fermions}

\author{R. J. Magyar}  \affiliation{Department  of  Physics,  Rutgers
University, 136 Frelinghuysen Road, Piscataway, NJ 08854-8019}

\author{K. Burke} 
\affiliation{Department of Chemistry
and Chemical Biology, Rutgers University, 610 Taylor Road, Piscataway,
NJ 08854-8019}


\begin{abstract}

A density  functional  theory  is developed  for
fermions in one dimension, interacting  via a  delta-function.
Such systems provide a natural testing ground for questions
of principle, as the local density approximation should work
well for short-ranged interactions.
The exact-exchange contribution to  the total energy
is a local functional of the  density.
A local density approximation for
correlation is obtained using perturbation theory
and Bethe-Ansatz results for the
one-dimensional  contact-interacting uniform Fermi  gas.
The  ground-state  energies are calculated  for two
finite systems, the analogs of Helium and of Hooke's atom.
The local approximation is shown to be excellent, as expected.

\end{abstract}


\date{\today} 

\maketitle


\section{Introduction}
\label{s:intro} 

Density functional theory (DFT) is a rigorous re-interpretation of the
quantum  many-body   problem  in  which  the   basic  object  uniquely
characterizing  a  system is  the  density,  $n(x)$,  rather than  the
many-body  wave-function.    This  view  is   particularly  suited  to
practical  calculations,  and DFT  has  been  applied successfully  to
solids  and molecules  for  quite some  time \cite{JG89}.   Researchers
typically focus  on Coulomb-interacting fermions  in three dimensions,
but  the  Hohenberg-Kohn  theorem,  or the  one-to-one  correspondence
between potentials and densities  \cite{HK64} upon which DFT is based,
holds for any  interaction and in any spatial  dimension.  We consider
the  contact or  delta-function  interaction between  fermions in  one
spatial dimension, 
\ben 
v_{ee}(x_i-x_j) = \lambda \delta(x_i-x_j),
\label{vint}
\een 
where  $x_i$ and $x_j$ ~represent ~the  ~spatial ~coordinates ~of
~the fermions, $\delta(x)$ is  the Dirac delta function, and $\lambda$
is the interaction strength.   The fermions have two spin states,
up and down.  The delta-function potential is a one-dimensional analog
to the Coulomb one as it scales in a similar fashion: 
$v_{ee}(ax) = v_{ee}(x) /a
$,   and   its   solutions   satisfy   the   energetically   important
particle-particle cusp condition  \cite{K73,D76}.  However, it differs
in  that  it  is   short-ranged.   There  is  no  simple
equivalence  between  $\lambda$  and  $e^2$,  the  Coulomb-interaction
strength,  although $\lambda$ can  be related  to a  scattering length
\cite{G97}.

One-dimensional model  interactions  are important for
several  reasons.    Perhaps  most  obviously,  they   are  useful  in
mathematical  and statistical physics\cite{T99,T86,K87}  to illustrate
problems  and   concepts  from  three-dimensional   physics  that  are
sometimes  hard to  conceptualize  due  to the  number  of degrees  of
freedom.    However,   our  primary   motivation   is   to  use   this
one-dimensional  model to  understand and  improve  density functional
theory.    Many    of   the    known   formal   properties    of   the
exchange-correlation  functional   are  true  in   this  case.   These
properties     include    behavior     under     uniform    coordinate
scaling\cite{LP85}, the  virial theorem,  and inequalities due  to the
variational  principle.   That properties  of  the  theory still  hold
should  prove  extremely useful  in  exploring time-dependent  density
functional theory\cite{RG84,MBAG02}, where formal properties are still
being explored.  Because  the interaction is not the  Coulomb one, the
unknown  exchange-correlation  functional  will,  of  course,  differ.
The  local density approximation  should be extremely
accurate  in this  case  because  of the  short-ranged  nature of  the
interaction.   Thus  in  our  case,  failures  of  the  local  density
approximation  (LDA)   can  be   ascribed  to  non-locality   that  is
independent of  the long-ranged nature of the  Coulomb repulsion.  Our
LDA could  be used  to study the  one-dimensional analog  of stretched
H$_2$ to identify whether  the proper description of dissociation into
individual atoms depends on  the long-ranged Coulomb interaction or is
due to  symmetry considerations alone.  
Another  interesting system on
which to  use one-dimensional DFT is the  one-dimensional solid.  This
delta-function interaction has already  been used to study problems in
DFT\cite{MWB1} but without the inclusion of any correlation effects, 
which are known to be important in one-dimensional systems.
Using DFT  to study  alternate interactions is  not new;  for example,
Capelle  and coworkers  have used  a similar  approach on  the Hubbard
model\cite{CLS3,LSOC3}.

It has been suggested that the delta-function model should give a good
representation of  the physics of one dimensional  fermions in certain
experimental  contexts  \cite{BH3,KF92,TCNT95,DHRBA97,QFYOA97}.  Since
one-dimensional systems are analytically, or at least computationally,
manageable, the  exact results are  useful to examine  situations when
standard techniques  fail.  For example,  the one-dimensional analogue
of  Helium  can be  examined  in detail  near  the  critical point  of
ionization when  the nuclear attraction and  interaction repulsion are
comparable.   This sort of  analysis is  demanding for  real systems,
and  finite-size  scaling and  infinite  dimensional approaches  are
necessary  \cite{K0,SK96,SK96c,SNK98}.   Carefully  understanding  how
systems  ionize and  how electronic  structure methods  reproduce this
critical phenomenon  are useful for  many chemical problems.   We will
examine this limit in detail in  future work and present only the most
basic results here.

Throughout, we assume that  our one-dimensional fermions have the same
mass as electrons, and we use atomic units ($e^2=\hbar=m_e=1$) so that
all energies are in Hartrees and all lengths in Bohr radii.


\section{Exact-Exchange Functional Density Functional Theory}
\label{s:exx}

In this section, we see  how the contact interaction affects the total
energy to first order in $\lambda$.  First order interaction theory is
traditionally  called the  Hartree-Fock approximation,  but  here, the
first order  interaction energy depends  explicitly on the  density so
that, for  this particular interaction, Hartree-Fock  is equivalent to
exact-exchange DFT.  Consequently,  exchange is treated exactly within
the local density approximation for this interaction.

According to the ~Hohenberg-Kohn theorem \cite{HK64}, the  ground-state  total
energy is a functional of the particle density:
\ben
E[\n] = T\s[\n] + U\H[\n] + E\xc[\n] + \int d x \; v\ext(x) 
\n(x) \nonumber\\
\label{etot}
\een
in a one-dimensional space where $U\H[\n]$ is the exactly known Hartree
or  classical density-density   interaction  contribution,  
$v\ext(x)$   is  the
given external   potential,  
$T\s[\n]$  is   the  exactly known kinetic   energy  of
non-interacting fermions at a given density, 
and $E\xc[\n]$ is the unknown exchange-correlation energy.   
The density is found by
studying the Kohn-Sham (KS) system,  the non-interacting counterpart to the
physical system \cite{KS65}.  
The Kohn-Sham equation is
\bea
\left(-\half \nabla^2 + v_{S,\sigma}([n_\uparrow,n_\downarrow];x)\right) \phi_{i,\sigma}(x)
= \epsilon_{i,\sigma} \phi_{i,\sigma}(x)
\label{ksequation}
\eea
where
$\phi_{i,\sigma}(x)$ is the $i^{th}$ KS orbital for spin-type, $\sigma$,
$\epsilon_{i,\sigma}$ is the KS eigenvalue,  
$v_{S,\sigma}(x)$ is the KS potential for spin-type, $\sigma$,
and $n_\uparrow$ and $n_\downarrow$ are the densities of up- and down-spins.
The Kohn-Sham potential is a functional derivative of the energy 
functionals,
\bea
v_{S,\sigma}(x) = v\ext(x) 
+ \frac{\delta U[n]}{\delta n_\sigma(x)}+ \frac{\delta E\xc[n]}{\delta n_\sigma(x)},
\label{vs}
\eea
where $n_\sigma(x)$ is the $\sigma$-spin density.
The spin-density is obtained from the occupied orbitals,
\bea
n_\sigma(x) = 
\sum_{i,\mbox{occ.}} 
|\phi_{i,\sigma}(x)|^2 .
\label{density}
\eea
Because  of the  anti-symmetry  of  the wave-function  under
particle  interchange, fermions with  like spins  do not
experience  the  contact interaction.
Only opposite spins interact directly.

The Hartree contribution depends only on the total particle density and 
is independent of how up and down fermions are distributed:
\bea
U_H[n] 
= \frac{\lambda}{2} \int dx\;dx'\; n(x) \delta(x-x') n(x') \nonumber \\
= \frac{\lambda}{2} \int dx\; n^2(x) .
\label{hartree} 
\eea
There is over-counting here because like spins do not interact, and
the exchange term must cancel these spurious like-spin interactions.
The exact-exchange term is
\bea
E_X[n_\uparrow,n_\downarrow] 
= -\frac{\lambda}{2} \int dx\; 
\left( n^2_\uparrow(x) + n^2_\downarrow(x) \right).
\label{ex}
\eea
Both terms can be derived using the standard rules 
of many-body perturbation theory \cite{M90} (see appendix A).

To simplify the notation,
let
\bea
\zeta(x) = \frac{n_\uparrow(x)-n_\downarrow(x)}{n_\downarrow(x)+n_\uparrow(x)}.
\label{zeta}
\eea
Then, the exact-exchange functional can be written,
\bea
E_X[n,\zeta] = -\frac{\lambda}{2} \int dx\; n(x)^2 (1 + \zeta(x)^2)/2,
\label{exx}
\eea
Note that for a one fermion system, 
we have $E_{X}[n]=-U_H[n]$ and contact-interacting
~exact-exchange is self-interaction free.  


\section{Local Density Correlation Functional from Deltium: 
The One-Dimensional Fermion Gas}
\label{s:deltium}

In order to obtain a local-density correlation functional, 
we review the  one-dimensional unpolarized Fermi gas, 
which we call deltium. 
This Fermi gas plays the role of the uniform electron gas in 
Coulomb-interacting DFT.
While the Coulomb-interacting Fermi-gas is a Fermi liquid, the 
one-dimensional delta-function interacting analog is a Luttinger liquid \cite{S93}.
The Hamiltonian is
\bea
\h= 
-\half \sum_i^N \frac{d^2}{dx_i^2} 
 +\lambda \sum_{i<j} \delta(x_i-x_j) .
\label{deltium}
\eea
The solution must be anti-symmetric under particle interchange 
and satisfy periodic boundary conditions on a ring of circumference, $L$.
This system has been examined previously\cite{G97,B97,NS83}.  
Because the wave-function is anti-symmetric under particle interchange, 
the fully polarized gas is not affected by the interaction.  
We consider the correlation in detail for only the fully unpolarized gas.  

The energy per particle of the non-interacting uniform gas is purely kinetic:
\ben
t (\n)= \frac{\pi^2}{24} \n^2 .
\label{ts1d}
\een
When interactions are present, the total energy per particle is 
\bea
\epsilon(\n)= t (\n) 
+ \epsilon_H(\n)
+ \epsilon\x(\n)
+ \epsilon\c(\n),
\label{energy2}
\eea
where 
$\epsilon\H(\n) = \lambda n / 2$ is the Hartree energy per particle,
$\epsilon\x(\n) = -\lambda n / 4$ is the exchange energy per particle,
and $\epsilon\c(\n)$ is the correlation energy per particle.
It is useful to define the following two terms.
Kinetic-like means that the energy per particle is proportional to $n^2$ 
like the non-interacting kinetic energy.
Hartree-like means that the energy per particle is proportional to $n$ 
like the Hartree energy. 

The ground-state ~energy per particle for deltium, Eq. (\ref{deltium}), 
can be 
found via Bethe Ansatz  methods \cite{LL63,G66,FL67,Y67,J68};
whereby, the uniform 
unpolarized
Fermi gas problem can be recast as a set of integral equations \cite{FB80}:
\bea
\tau(y) = \frac{1}{2\pi} 
+  \frac{2}{\pi}
\int_{-\infty}^\infty d\Lambda 
\frac{\lambda\sigma(\Lambda)}{\lambda^2 + 4(y-\Lambda)^2}
\label{barho}
\eea
and
\ben
\sigma(\Lambda) = \frac{1}{2\lambda} 
\int_{-k_{max}}^{k_{max}} dy\; 
\sech\left(\pi (y-\Lambda) /\lambda \right) \;\tau(y)
\label{basigma}
\een
where $\tau$ is the number of occupied states per wave-vector label, $y$; 
and  $\sigma$ is  the number of occupied down-spin states per a different 
wave-vector label, $\Lambda$.
In the high-density limit, $k_{max} = \pi n /2$,
and in the low-density limit, $k_{max} = \pi n$.
Equations (\ref{barho}) and (\ref{basigma}) 
must be solved self-consistently for $\tau$ and $\sigma$ 
at a chosen value of $k_{max}$ to obtain
the  ground-state  energy.
In order to do this,
the integrals are transformed  to  the  interval, ~[-1,1],  
and integrated using six-point quadrature rules with 400 mesh points.    
The density is
\ben
n = \int_{-k_{max}}^{k_{max}} dy\; \tau(y) ,
\een
and the total energy per particle is
\ben
\epsilon = \frac{1}{2n} \int_{-k_{max}}^{k_{max}} dy\; y^2 \; \tau(y) .
\een
The correlation energy per particle for
a  wide  range   of  densities can,  in   the  spirit  of 
three-dimensional DFT, 
be parameterized for practical calculations.  
We consider
both  the high and  low-density  limits analytically  and numerically.
Since we are concerned with parameterizing correlation energy, we
subtract out the known  kinetic, Hartree, and exchange contributions.  


The low-density limit is the large $\lambda$ limit.  In this
limit, the  opposite-spin Fermions repel  each other so  strongly that
the   interaction  mimics   Fermi   anti-symmetrization.   Thus,   the
interaction energy per particle  is kinetic-like.  This means that the
first term for the correlation energy must cancel the Hartree energy, and
the next term in the correlation energy  must be kinetic-like.
The correlation energy per particle  is 
\bea
\epsilon\c (n) = -\frac{\lambda}{4} n 
+b_1  n^2 
-\frac{b_2}{\lambda} n^3 + {\cal O}\left(\frac{n^4}{\lambda^2}\right) \nonumber\\ 
\label{lowdens}
\eea 
with $b_1=\pi^2/8$
and $b_2=4.560971$.
The first terms  in Eq. (\ref{lowdens})
exactly cancels  the Hartree and exchange energies, and the next, $b_1$, 
provides 
the kinetic-like contribution to the energy.   
We determine the coefficient, $b_2$, from numerical analysis of the 
Bethe Ansatz results at $\lambda=1$.
Specifically, we subtract the known contributions from $\epsilon(n)$, 
divide by $n^3$,  
and plot the result as a function of $n$.  
The extrapolated intercept at $n=0$ is $b_2$.

In ~the ~high-density ~limit, the ~interaction ~is ~perturbation-like
and the correlation energy per particle approaches a constant:
\bea 
\epsilon\c (n) = 
-c_1 \lambda^2  +  c_2 \lambda^3 / n + {\cal O}\left(\frac{\lambda^4}{n^2}\right) 
\label{highdens}
\eea 
with  $c_1=1/ 24 $ and $c_2=0.006151$.  
The first term, $c_1$ is found in appendix A using perturbation theory.  
We determine $c_2$ from numerical analysis of the Bethe Ansatz results.
We find the expansion coefficients
by subtracting the known contributions from $\epsilon(n)$ and plotting the remainder 
as a function of $1/n$.  The coefficients are the extrapolated 
intercept and slope as a function of $1/n$.
The correlation energy per 
particle of the uniform gas approaches a finite value as $n\rightarrow\infty$ 
in contrast to jellium, because the contact interaction is short-ranged.

\begin{figure}[t]
\begin{center}
\unitlength1cm
\begin{picture}(6.5,7.0) 
\put(0.5,0.5) 
{\psfig{figure=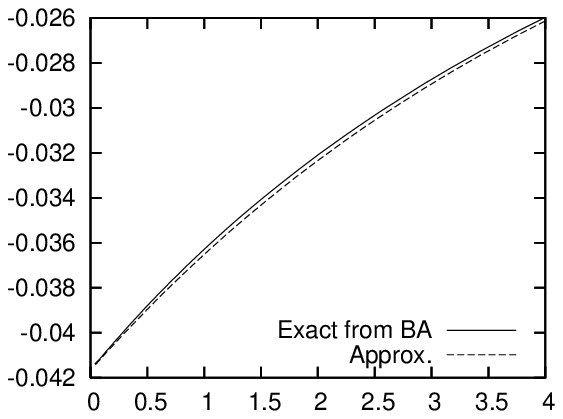,width=8cm,height=7cm}}
\setbox6=\hbox{\large $\epsilon\c(\n)$}
\put(0.2,4.0){\makebox(0,0){\rotl 6}}
\put(4.5,0.0){\large $1/\n$}
\end{picture}
\caption{Correlation energy per particle for Deltium, 
the one-dimensional uniform contact-interacting ($\lambda=1$) 
Fermi gas,
in the high-density limit.
The solid line is the exact result 
calculated from the solutions of the Bethe-Ansatz integral equations.
The long-dashed line is the simple 
LDA parameterization given by Eq. (\ref{eclda}).}
\label{f:unif1} 
\end{center}
\end{figure}

\begin{figure}[t]
\begin{center}
\unitlength1cm
\begin{picture}(6.5,7.0) 
\put(0.5,0.5) 
{\psfig{figure=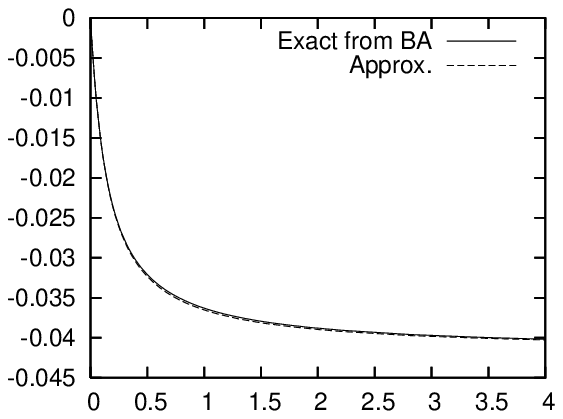,width=8cm,height=7cm}}
\setbox6=\hbox{\large $\epsilon\c(\n)$}
\put(0.2,4.0){\makebox(0,0){\rotl 6}}
\put(4.5,0.0){\large $\n$}
\end{picture}
\caption{The same as Fig \ref{f:unif1} but for the low-density limit.}
\label{f:unif2} 
\end{center}
\end{figure}

A [2,2] Pad{\'e} parameterization of  the correlation energy per particle
is 
\bea 
\epsilon\c^{LDA}(n) \approx 
\left(
\frac{a n^2 + b n + c}
{n^2+d n + e } 
\right)   
\label{epsclda1}
\eea 
with 
$a=-0.0416667$,
$b=0.004475$,
$c=0$,
$d=0.254998$,
and
$e=0.017900$.
Note that $c$ is zero because the correlation 
energy per particle vanishes in the low-density limit.
This  approximation
gives  the first and second terms of both the high and low-density limits correctly.
The  parameterization 
of $\epsilon\c$, the correlation energy per particle, has a maximum
error of 0.7$\%$ and is highly accurate for the important high density regions.  
As shown in Figures \ref{f:unif1} and \ref{f:unif2}, the parameterization 
is almost indistinguishable from the exact numerical result.

For the fully polarized case, the interaction does not contribute and 
\bea
E_{C}^{LDA} [n] = 0 \label{ecldapol}.
\eea
We can combine these results and construct a local-density correlation energy functional:
\bea
E_{C}^{LDA} [n] =
\int dx\; n(x) \;\epsilon\xc^{unif.}( n(x)) f(\zeta(x)) \\
\approx  
-\int  dx\; \left(
\frac{a n(x)^2 + b n(x)}
{n(x)^2+d n(x) + e } 
\right)  f(\zeta(x)).   
\nonumber\\ \label{eclda}.
\eea
A simple suggestion for $f(\zeta)$ which gives both polarized and unpolarized limits exactly is
\ben
f(\zeta) = 1-\zeta^2 
\label{simplef}
\een
which is the $\zeta$ dependence of $U_{H}$ plus $E\x$.

%
%


\section{Diracium, the Delta-Function Interacting Analog of Helium}
\label{s:diracium}
\begin{table}[t]
\begin{center}
\begin{tabular}{|lcccc|} 
\hline
Z          & Exact        & $\Delta$LDA(mH)     & $\Delta$ EXX &  $\Delta$ Pert. Th.\\  	
$Z_{crit}$ & -0.070276	  & -45    	    & -70	  &  46 \\
0.5        & -0.129281	  & -11             & -129	  &  34 \\
1          & -0.647225	  & -3.3	    & -64	  &  16 \\
2          & -3.155390	  & -1.0	    & -72	  &  7.4  \\
4          & -14.159190   &  0.7	    & -76	  &  3.6  \\
8          & -60.161010	  &  1.8	    & -78	  &  1.7  \\
100        & -9950.1630	  &  3.2	    & -80         & -0.2  \\
\hline
\end{tabular}
\caption{\label{t:resdirac} Total ground-state 
energy for Diracium and errors (in milliHartrees)
of various approximations with $\lambda=1$. 
The exact results are from a numerical solution of the problem as outlined 
by Ref. \cite{R71},
the second-order perturbation values are also given in the same reference.  
The EXX  and the LDA results are from a self-consistent 
solution using the exact-exchange functional, Eq. (\ref{ex}), 
and the  LDA functional of Eq. (\ref{eclda}) respectively.
$Z_{crit}=0.377115$ is the critical value at which the system ionizes \cite{R71}.
}
\end{center}
\end{table}
%
\begin{figure}[t]
\unitlength1cm
\begin{picture}(6.5,7.0) 
\put(0.5,0.5) 
{\psfig{figure=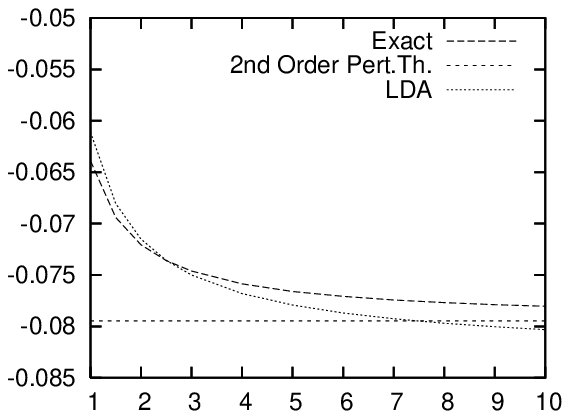,width=8cm,height=7cm}}
\setbox6=\hbox{\large $E_{C}$}
\put(0.2,4.0){\makebox(0,0){\rotl 6}}
\put(4.5,0.0){\large $Z$}
\end{picture}
\caption{Correlation energy for Diracium, the one-dimensional analog of Helium 
($\lambda=1$).
The solid line  is the exact result, the short-dashed  line is the LDA
result,  and the  long-dashed line  is the  second  order perturbation
theory  result.    Note  that   we  extract  the   correlation  energy
approximately by subtracting  the self-consistent exact-exchange total
energy from the  exact total energy.  The exact-exchange
density becomes  an unreliable  approximation to the  exact-density as
$Z$ decreases and  at $Z=1/2$ the density is  even qualitatively wrong
as  the  self-consistent  exact-exchange  density  no  longer  exists.
Therefore,  we  terminate  the  plot  at $Z=1$  where  we  expect  the
self-consistent   exact-exchange    density   to   be    an   accurate
representation of the exact density.}
\label{f:ec} 
\end{figure}

In order to assess the usefulness of this local approximation 
to handle one-dimensional problems, we start with perhaps 
the most difficult test case, a completely non-uniform system, diracium.
This is the one-dimensional analog of Helium with the traditional Coulomb
terms replaced by delta-functions.
The system is described by the Hamiltonian:
\bea
\h= 
-\half  \frac{d^2}{dx_1^2}  - \half  \frac{d^2}{dx_2^2} 
- Z \delta(x_1)-  Z\delta(x_2) +\lambda \delta(x_1-x_2)
\nonumber \\
\label{diracium}
\eea
and the eigenvalue equation,
\bea
\h
\Psi_{\sigma\sigma'}(x_1,x_2)
= E
\Psi_{\sigma\sigma'}(x_1,x_2) .
\label{hpsi}
\eea
where $x_1$ and $x_2$ are the positions of the fermions,
$\sigma$ and $\sigma'$ are the spin-like labels of the fermions,
$Z$ is the magnitude of the external potential,
and $\Psi$ is an antisymmetric Fermi wave-function 
which vanishes as $x\rightarrow\infty$.
The ground-state is a spin-singlet.

First, we solve the
model analytically within the  exact-exchange approximation.  Then, we
introduce our  local density approximation to  the exchange-correlation
energy.  Finally, we present the exact energy eigenvalues.

With the exact-exchange functional, Eq. (\ref{exx}), and no functional for correlation, 
the Kohn-sham single-orbital equation is 
\bea
-\half \frac{d^2}{d^2x}\phi(x) - Z \delta(x) \phi(x) 
+ \lambda |\phi^2(x)|^2\phi(x) = -\epsilon_{KS} \phi(x).
\nonumber \\
\label{exxks}
\eea
Equation (\ref{exxks}) can be solved using elementary techniques.
The resulting eigenvalue is
\bea
\epsilon_{KS} = -\half \left(Z-\half\lambda\right)^2 ,
\label{exxeig}
\eea
and the Kohn-sham orbital \cite{LSY92} is
\bea
\phi(x) = \frac{1}{\sqrt{\lambda}}\left(Z-\half\lambda\right)
\csch\left[\left(Z-\half\lambda\right) |x| 
+ \alpha 
\right] \nonumber \\
\label{exxkswf}
\eea
with
\bea
\alpha = \arccoth\left(\frac{Z}{Z-\half\lambda}\right).
\label{alpha}
\eea
This is unbound at $Z<\half\lambda$.
The calculated total energy is not just the sum of the KS eigenvalues, rather it is 
\bea
E[\n] = \sum_{occ.} \epsilon_{KS} + U[\n] + E_{XC}[\n] \nonumber\\
- \int dx \; v_{H}(x) n(x)
- \int dx \; v_{XC}(x) n(x),
\label{exxetot}
\eea
or explicitly,
\bea
E_{EXX} =  -Z^2 + \frac{Z\lambda}{2} - \frac{\lambda^2}{12} .
\label{ehf}
\eea


Next, we solve the KS equation using the local density approximation to the correlation.
The LDA KS equation,
\bea
-\half \frac{d^2}{d^2x}\phi(x) - Z \delta(x) \phi(x) 
+ \lambda |\phi^2(x)|^2\phi(x)  \nonumber \\
+v_C(x) \phi(x)
= - \epsilon_{KS} \phi(x)
\label{ldaks}
\eea
with
\bea
v_C(x) = 
\left( \frac{1}{n(x)}
+ \frac{a}{a n(x)+b} -\frac{2 n(x)+d}{n(x)^2+d n(x)+e} \right) \epsilon_C(x)
\nonumber\\
\label{vclda}
\eea
Equation (\ref{ldaks}) 
is 
solved numerically via a self-consistency cycle and the shooting method.

The exact ground-state energies were obtained previously by Rosenthal \cite{R71} by
transforming  to  momentum  space  and  reducing the  problem  to  the
solution  of a one-dimensional integral  equation. 
While  this method
converges  quickly to  the exact  energy  eigenvalue, it  is not  well
suited to  give real space  wave-functions and densities.   Instead, we
take the calculated eigenvalue, $E$, 
as input and reduce the eigenvalue problem,
Eq. (\ref{diracium}), 
to
a  differential  equation.  The differential equation 
can then converted to an  integral equation 
using Green's function techniques:
\bea
\Psi(x,y) = 
\frac{Z}{\pi} \int_{-\infty}^{\infty} dx'\;      
K_0 \left( \sqrt{-2 E} \sqrt{ (x-x')^2 + y^2} \right) \Psi(x',0) \nonumber \\ 
+\frac{Z}{\pi} \int_{-\infty}^{\infty} dx'\;      
K_0 \left( \sqrt{-2 E} \sqrt{ x^2 + (y-x')^2} \right) \Psi(x',0)  \nonumber \\ 
-\frac{\lambda}{\pi} \int_{-\infty}^{\infty} dx'\;
K_0 \left( \sqrt{-2E} \sqrt{ (x-x')^2 + (y-x')^2 } \right) \Psi(x',x') , \nonumber \\ 
\label{getpsi}
\eea
where $E$ is the ground state energy and 
$K_0$ is the zeroth-order modified Bessel-function and is the Green's function 
for this particular two-dimensional equation.  
By setting $y$ equal to $x$ and then $0$, 
we can arrive at a set of two coupled integral equations which can 
be solved self-consistently.  Note that although the Bessel function has a divergence, 
the integral is finite.  Once  $\Psi(x,0)$ and $\Psi(x,x)$  are known, we 
can construct the wave-function at any point in space. If the inputed $E$ is exact, 
then solution of this equation yields the exact ground-state wave-function.

\begin{figure}[t]
\unitlength1cm
\begin{picture}(6.5,7.0) 
\put(0.5,0.5) 
{\psfig{figure=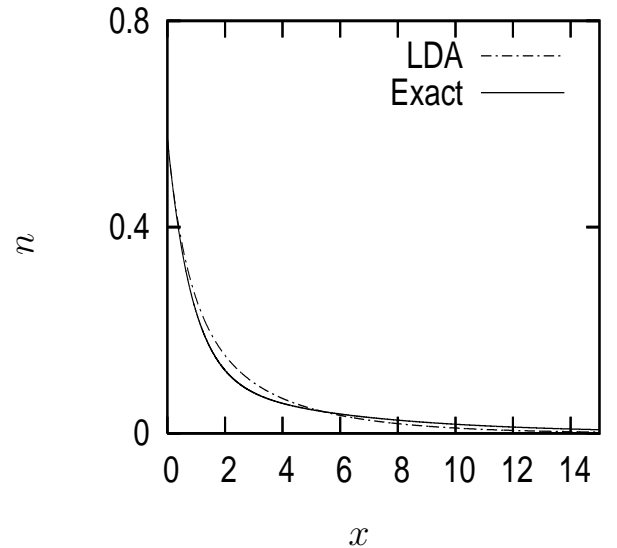,width=8cm,height=7cm}}
\setbox6=\hbox{\Large $n$}
\put(0.2,4.0){\makebox(0,0){\rotl 6}}
\put(4.5,0.0){\Large $x$}
\end{picture}
\caption{Comparison of the self-consistent LDA density (dashed line)
 with the exact (solid line)  for $\lambda=1$ and $Z=0.5$.  
This is an extreme case where the exact-exchange no longer binds.}
\label{f:dens} 
\end{figure}

In  Table  \ref{t:resdirac}, we  see  that LDA is  greatly  more
accurate than the EXX functional.  Second order perturbation theory is
only  more   accurate for $Z\geq 7$,
but  that  is  at   a  much  larger
computational  cost  since  the  second  order  contribution  requires
calculation of  the entire spectrum  of unoccupied orbitals.
LDA remains bound and 
gives a reasonable result (within a factor of three) even 
at the critical potential strength, $Z_{crit}$.   In Figure \ref{f:dens}, 
we see that the LDA gives a qualitatively correct density for $Z=1/2$ 
where the EXX result is no longer even bound.

\begin{figure}[t]
\unitlength1cm
\begin{picture}(6.5,7.0) 
\put(0.5,0.5) 
{\psfig{figure=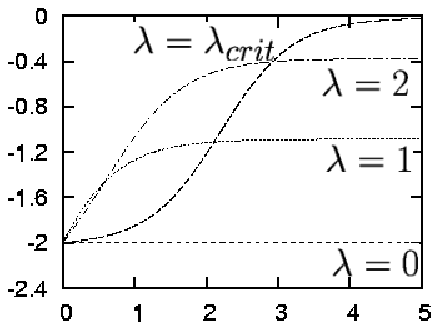,width=8cm,height=7cm}}
\setbox6=\hbox{\Large $\frac{d\ln n}{dx}$}
\put(0.2,4.0){\makebox(0,0){\rotl 6}}
\put(4.5,0.0){\Large $x$}
\end{picture}
\caption{The behavior of the density for Diracium ($Z=1$) 
at various interaction strengths, $\lambda$.  
We plot $d \ln n(x)/dx$ to high-light the asymptotic behavior of the density.
For $\lambda>\lambda_{crit}$, the system is ionized.}
\label{f:kais} 
\end{figure}

There exist standard theorems about the decay of the density 
away from the attractive nuclear potential in the 
three-dimensional Coulomb interacting case \cite{HL95}.  For example,
\ben
n(r) \rightarrow e^{-2\alpha r}
\een
as $r\rightarrow\infty$ where $\alpha  = \sqrt{2  I}$.   
This theorem also holds for our
one-dimensional model with $r=|x|$.  There is much interest  in critical values
of $Z$  at which an  atom can no  longer bind its  outermost electron\cite{K0,SK96,SK96c}.
Understanding this limit yields  information on  the existence of  negative ions.   
A most
interesting question  is: 
As $Z\rightarrow  Z_{crit}$ and $I\rightarrow
0$,  how  does  the density  decay?  We  study  this directly  in  our
one-dimensional  example by varying $\lambda$ keeping $Z$ fixed at 1.  
For large enough $\lambda$, the system will ionize.
Note that $\lambda_{crit}=1/Z_{crit}$.    Figure  \ref{f:kais}  shows  $ d \ln n/dx$  as
$\lambda\rightarrow 1/Z_{crit}$ for $Z=1$. 

The ~second-order ~perturbation theory result 
for Diracium differs from the high density LDA result.  
The second-order perturbation theory result is:
\ben
E_C = (- 3/8 + 2/3\pi +1/12) \lambda^2
= -0.0795 \lambda^2.
\een
The first two terms are the exact contribution to the 
total energy to order $\lambda^2$ \cite{WS70}.  From this, we subtract 
the final term which is the exchange contribution via the self-consistent density. 
The high-density LDA correlation result is
\ben
E_C = -\lambda^2/12 = -0.083333 \lambda^2
\een
Because the interaction is short-ranged, LDA correctly scales to a constant 
in the high-density limit, and in contrast to three-dimensional Coulomb 
DFT \cite{GL93} is highly accurate.

\begin{figure}[t]
\unitlength1cm
\begin{picture}(6.5,7.0) 
\put(0.5,0.5) 
{\psfig{figure=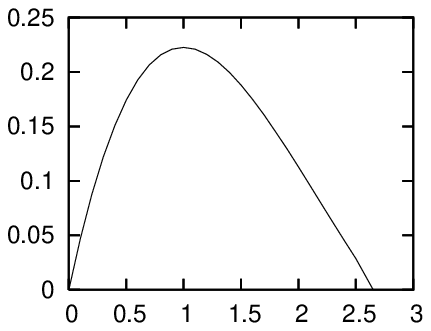,width=8cm,height=7cm}}
\setbox6=\hbox{\large $\langle v_{ee}\rangle$}
\put(0.2,4.0){\makebox(0,0){\rotl 6}}
\put(4.5,0.0){\large $\lambda$}
\end{picture}
\caption{
The expectation value of the interaction 
at various interaction strengths, $\lambda$.  
Beyond  $\lambda=\lambda_{crit} (2.6517)$, the system is ionized 
and the interaction-energy vanishes.}
\label{f:vee} 
\end{figure}

Another  ~interesting  ~quantity to  consider  as the  particle-particle
interaction  grows   stronger  is  the  interaction   energy  or  the
expectation value of $\hat{v}_{ee}(\hat{x}_i-\hat{x}_j)$.
For small $\lambda$, the interaction
energy grows in magnitude as $\lambda$ grows, 
but at  $\lambda \approx 0.9$,
this trend reverses.   For  $\lambda \geq \lambda_{crit}$, the
system is ionized, so that the interaction energy is zero.
Figure  \ref{f:vee} shows that, as $\lambda\to\lambda_{crit}$ from
below, the approach to this discontinuity is linear.
This information is valuable in studying the approach to ionization,
and may also be true for real two-electron ions.


%
%
%

\section{Delta-Function Interacting Hooke's Atom}

Another test of this one-dimensional LDA is the analog to Hooke's atom \cite{T93},
\bea
\h= 
-\half  \frac{d^2}{dx_1^2}  - \half  \frac{d^2}{dx_2^2} 
+\half \omega^2 x_1^2+  \half \omega^2 x_2^2 +\lambda \delta(x_1-x_2) \;
\nonumber \\
\eea
where $\omega$ determines the strength of the harmonic well potential.
This model has been used to model one-dimensional quantum dots \cite{BH3}, 
and its excitations have been studied using time-dependent DFT \cite{MZCB3}. 
The exact wave-function is given in terms of Whittaker functions and confluent 
hyper-geometric functions \cite{BH3}.  The total energy is
\ben
E = \half \omega +\epsilon
\label{etothooke}
\een
with $\epsilon$ obtained from the solution of \cite{BH3,MZCB3}
\ben
2\sqrt{2\omega} \;
\Gamma\left(-\frac{\epsilon}{2\omega}+\frac{3}{4}\right)\; / \;
\Gamma\left(-\frac{\epsilon}{2\omega}+\frac{2}{4}\right) = -\lambda .
\label{epsilon}
\een
In Table \ref{t:reshooke},  we see that LDA greatly  improves over the
exact exchange  formalism for all  values of $\omega$.  

In the high-density or weak-coupling ($\lambda\to 0$) limit,
this ~system ~behaves similarly to diracium, described above.
The total energy can be described perturbatively
\ben
E=\omega+\lambda c_1 \omega^\half + \lambda^2  c_2  + \dots
\een
where $c_1=1/\sqrt{2\pi}=0.399$ and $c_2=\left(\gamma +
\psi^{(0)}(\half)\right)/ 4 \pi=-0.110318$ with
$\gamma$ being the Euler constant and $\psi^{(0)}$,  
the zeroth-order polygamma function.
The $\lambda^2$-term above is the high-density limit of the correlation
energy  plus   an  exchange   contribution.   In  DFT,   the  exchange
contribution  is  the  first-order  contribution  in  $\tilde{\lambda}$  where
$\tilde{\lambda}$  is the  coupling  constant for  a  fixed fermion  density.
Here, $\lambda$  is the  interaction strength, and  when it  varies so
does the ground-state density.   This means that in a self--consistent
calculation, exchange will contribute  to second order in $\lambda$ to
the total energy.   We saw this in Eq.  (\ref{ehf}) for diracium.  The
same is  true here.  Like for diracium,  LDA is 
very  accurate in the high-density 
limit.  In
order  to properly  compared the  the LDA  correlation energy  with the
exact one, we  need to extract the exact density.

Since Hooke's atom stays bound for arbitrarily weak  $\omega$,
we can see how well our LDA describes
the low-density limit.
The  low-density  limit  is  particularly  challenging  for Coulomb-interacting
functionals \cite{SZRS78} because of the strong correlation.  LDA and
generalized gradient approximations (GGAs)
behave qualitatively correctly but often err by as  much as a factor
of two for the exchange-correlation  energy.
For Hooke's atom, the exact result for the total energy in the strong-coupling limit is
\ben
E = 2 \omega.
\een
The low density limit is related to the large coupling constant limit.
As $\lambda\to\infty$, the lowest energy solution in the relative
coordinate, $|x_1-x_2|$,
is simply the first non-interacting 
excited state because inserting a
node at $x_1=x_2$ minimizes the interaction energy.
We find that LDA is
remarkably accurate  in this regime, 
so the errors in Coulomb-interacting DFT
can be ascribed to the long-range interaction.

The  energy in
this  regime becomes kinetic-like.   The exact-exchange  functional is
Hartree-like and will fail to  capture the proper energetics.  The LDA
however  cancels  the   Hartree-like  exchange  contributions  and  is
kinetic-like.   In the  low-density  regime the  density  is close  to
uniform locally, so we expect the energy per particle to be similar to
low-density deltium.  This is  in fact the  case and is reflected  by the
high accuracy of the LDA in the low density regime.

\begin{table}[t]
\begin{center}
\begin{tabular}{|lccc|} 
\hline
$\omega$ & Exact      & $\Delta$ EXX (mH)& $\Delta$LDA \\  
0.001    & 0.001950   &	- 6.1	 & -0.08   \\
0.01     & 0.018510   &	-12.6	 & -0.7   \\
0.1      & 0.161410   &	-48.2	 & -3.5   \\
1        & 1.306750   &	-72.2	 & -6.3   \\
10       & 11.157330  &	-82.9	 & -6.4   \\
100      & 103.881057 & -86.5	 & -5.7   \\
\hline
\end{tabular}
\caption{\label{t:reshooke}  
Total ground-state 
energy for 
 the contact-interacting 
Hooke's Atom and errors (in milliHartrees) of various approximations with $\lambda=1$. 
Exact is from a numerical solution of Eqs. \ref{etothooke} and \ref{epsilon}.  
EXX is exact-exchange.  
LDA is according to the parameterization Eq. \ref{eclda}.}
\end{center}
\end{table}



\section{conclusion}
\label{s:conclusions}

In this paper, we examined a one-dimensional density functional theory
of contact  interacting fermions.  We noted that  exact-exchange is an
explicit   density   functional    and   developed   a   local-density
approximation  for  correlation.  We  have  applied these  functionals
successfully to  two simple models demonstrating the  high accuracy of
LDA here.  Although  LDA is highly accurate in these  cases, it is not
exact.    This  result   is   consistent  with   the  observation   in
Ref.  \cite{BPL94} that  LDA is  not  exact in  the short  wave-length
limit.


This model interaction  and LDA can be used  to illustrate and explore
problems in  DFT.  Examples include ground-state  symmetry problems in
stretched H$_2$  and the interacting-fermion  one-dimensional solid (a
generalized  Kronig-Penney  model)  as  a  model  band-gap  problem.
Excited-state  Bethe-Ansatz results might  be used  to derive  a local
current-density functional for  this interaction.  This delta-function
interaction  has  already been  used  in  scattering  problems and  in
pedagogy, and  we hope  that our local-density  correlation functional
finds fruitful applications in these areas as well.


\section{Acknowledgments}

We would like to thank C. Rosenthal for discussions.
This work was supported by the National Science Foundation 
under grant number CHE-9875091 and the Department of Energy
under grant number DE-FG02-01ER45928.


\appendix
  \renewcommand{\theequation}{A-\arabic{equation}}
  \setcounter{equation}{0}  
  \section*{APPENDIX A}  


Here, we  find the  high density limit  of the correlation  energy per
particle for  deltium using  perturbation theory and  the diagrammatic
approach in  momentum space \cite{M76}.  
The Fourier  transform of the
interaction potential  is 
\bea 
V(q)  = \frac{\lambda}{L} \int_{-L}^{L}
dx\; \delta(x) e^{i q x}  = \frac{\lambda}{L}  
\label{veefourier}
\eea 
where $L$ is the length between the periodic boundaries.
As noted earlier,
like  spin fermions  will not  interact via  the  delta-function, this
means that only vertices that  connect opposite spins will enter into
the  perturbative series.   This  is a tremendous simplification as  many
diagrams will cancel.  A further  simplification is that the interaction
is independent of the momentum transfer, $q$.
%
\begin{figure}[b]
\unitlength1cm
\begin{picture}(6.5,2.5) 
\put(0.0,0.0) 
{\psfig{figure=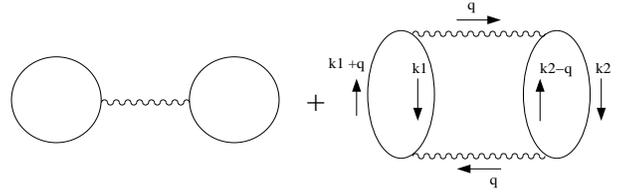,width=8cm,height=2.5cm}}
\end{picture}
\caption{
First and second order contributions to the interaction energy.
Spin labels are omitted but the two loops in each diagram must have opposite spins.
$q$ is the momentum transfer,
$k_1$ and $k_2$ label particle momenta, and 
$k_1+q$ and $k_2-q$ label hole momenta.}
\label{f:feyn} 
\end{figure}

To  first  order, only  the  Hartree  diagram  between opposite  spins
contributes to the total energy.   
This  is  the  first diagram  in  Figure  \ref{f:feyn}.
Evaluation does not requires integration over internal momentum.  
The loop integrals have pre-factors of $L^2/ 4 \pi^2$, and
integration over the loops results in a factor of $2 k_F = \pi n$ each, 
where
$k_F = \pi /2 n$ is the Fermi momentum of non-interacting deltium.
The
symmetry factor  of $1/2$ is canceled  by a sum over the two possible
pairs of spin.  The final  energy per particle is obtained by dividing
by  $N$, the  total particle number.  We find  
\bea 
N \epsilon_{HX}  =
\frac{\lambda}{L} \left(\frac{L^2}{4\pi^2}\right) 
\pi^2 n^2 \;
\rightarrow \;
\epsilon_{HX}  =
\lambda \frac{n}{4} .  
\label{ehxhd1}
\eea

To  second order,  only one  more  diagram contributes.   This is  the
two-bubble  diagram shown  second  in Figure  \ref{f:feyn}.  From  the
standard rules  of perturbation theory, 
\bea 
N  \epsilon_C = \nonumber \\  
-\frac{\lambda^2}{L^2}  \frac{L^3}{8\pi^3}  \int_{-\infty}^\infty
dq\; \int_{-k_f}^{k_f} dk_1\;  \int_{-k_f}^{k_f} dk_2\; \frac{1}{q ( q
+ k_1-k_2)}  \label{echd1}
\nonumber \\  
\eea 
with $|k_1+q|>1$  and $|k_2-q|>1$,
$k_1$ and  $k_2$ are particle  momenta, and $q$ is the  momentum transfer.
The limits of integration and constraint inequalities ensure that 
particle  have less momentum than the Fermi-momentum, and
holes have higher momentum than the Fermi momentum.
Once again, the sum over the two possible spin-arrangements cancels the symmetry factor of $1/2$.
To solve Eq. (\ref{echd1}) exactly, we rescale as follows: 
$q   = k_F x$, 
$k_1 = k_F y$, 
and
$k_2 = k_F z$.   After some algebra, we find the correlation energy per particle: 
\bea
\epsilon_C =  -\frac{\lambda^2}{8\pi^3}\left(\frac{L}{N}\right) 
\left(\frac{\pi n}{2}\right)\; {\cal I} = -\frac{\lambda^2}{24}
\label{echd2}
\eea
using the quadrature result,
\bea
{\cal I} =
2 \int_2^\infty dx\;
  \int_{-1}^1 dy\;
  \int_{-1}^1 dz\;
  \frac{1}{x (x+y-z)} \nonumber\\
+
2 \int_0^2 dx\;
  \int_{1-x}^1 dy\;
  \int_{-1}^{-1+x} dz\;
  \frac{1}{x (x+y-z)}
= \frac{2\pi^2}{3}. \nonumber\\
\label{quadrature}
\eea



\end{document}